\documentclass{emulateapj}
\usepackage{graphicx}
\slugcomment{Ver 1, 3 Feb 2011}

\shorttitle{CO Einstein Ring SMG MM1842}
\shortauthors{Lestrade, et al.}

\begin{document}

%\input{defsv1}
%%%%%%%%%%%%%%%%%%%%%%%%%%%

%\title{A Molecular Einstein Ring Towards the  z=3.93 Submillimeter Galaxy MM18423+5938}

\title{A MOLECULAR EINSTEIN RING TOWARD THE z=3.93 SUBMILLIMETER GALAXY MM18423+5938}

%%%%%%%%%%%%%%%%%%%%%%%%%%%
%% \author{Jean-Francois Lestrade \altaffilmark{1}, Chris L. Carilli\altaffilmark{2}, Jean-Paul Kneib\altaffilmark{3}, Karun Thanjavur \altaffilmark{4,5}}
\author{Jean-Fran\c cois Lestrade \altaffilmark{1}, Chris L. Carilli\altaffilmark{2}, Karun Thanjavur\altaffilmark{3,4}, 
Jean-Paul Kneib\altaffilmark{5}, Dominik A. Riechers\altaffilmark{6},
Frank Bertoldi\altaffilmark{7}, Fabian Walter\altaffilmark{8}, Alain Omont\altaffilmark{9}}

%%%%%%%%%%%%%%%%%%%%%%%%%%%

 \altaffiltext{1}{Observatoire de Paris, CNRS, 61 Av. de l'Observatoire, F-75014, Paris, France jean-francois.lestrade@obspm.fr}
 \altaffiltext{2}{NRAO, Pete V. Domenici Array Science Center, PO Box O, Socorro, NM 87801, USA} 
 \altaffiltext{3}{Canada France Hawaii Telescope Corporation, HI 96743, USA.}
 \altaffiltext{4}{Department of Physics \& Astronomy, University of Victoria, 
         Victoria, BC, V8P 1A1, Canada.}
\altaffiltext{5}{Laboratoire d'Astrophysique de Marseille, Observatoire d'Astronomie Marseille-Provence, BP 8, F13376 Marseille, 
France}
\altaffiltext{6}{Astronomy Department, Caltech, 1200 East California boulevard, Pasadena,  CA 91125, USA}
\altaffiltext{7}{Argelander Institute for Astronomy, University of Bonn, Auf dem Hugel 71, 53121 Bonn, Germany}
\altaffiltext{8}{Max-Planck Institute for Astronomy, Konigstuhl 17, D-69117  Heidelberg, Germany}
\altaffiltext{9}{Institut d'Astrophysique de Paris, 98bis Bld Arago, F75014, Paris, France}

%%%%%%%%%%%%%%%%%%%%%%%%%%%

\begin{abstract}

We present high resolution imaging of the low order ($J=1$ and 2) CO line emission
from the $z = 3.93$ submillimeter  galaxy (SMG) MM18423+5938 using the Expanded
Very Large Array, and optical and near-IR imaging using the Canada-France-Hawaii Telescope.  
This SMG with a spectroscopic redshift was thought to be gravitationally lensed
given its enormous apparent brightness. We find that the CO emission is consistent with 
a complete Einstein ring with a major axis diameter of $\sim 1.4''$, indicative of lensing.  
We have  also identified the lensing galaxy as a very red elliptical coincident 
with the geometric center of the ring and estimated its photometric redshift $z\sim1.1$.
A first estimate of the lens magnification factor is $m \sim
12$. The luminosity $\rm L'_{CO(1-0)}$ of the $\rm CO(1-0)$ emission is $2.71\pm 0.38 \times 10^{11}~m^{-1} $~K km s$^{-1}$ pc$^2$,
and, adopting the commonly used conversion factor for ULIRGs, the molecular gas mass  is 
${\rm{M(H_2)}} =  2.2 \times 10^{11}~m^{-1} ~{\rm M_\odot}$, 
 comparable to  unlensed SMGs if corrected by $m \sim 12$.   
Our revised estimate of the far-IR luminosity  of MM18423+5938 is  
$2 \times 10^{13} ~m^{-1} < L_{\rm FIR} < 3 \times 10^{14} ~m^{-1} ~{\rm L_{\odot}}$, 
comparable to  that of ULIRGs. Further observations are required to quantify 
the star formation rate in MM18423+5938 and to constrain the mass model of the lens in more detail.  

\end{abstract}

%%%%%%%%%%%%%%%%%%%%%%%%%%%


%%%%%%%%%%%%%%%%%%%%%%%%%%%
\section{Introduction} \label{intro}
There is mounting evidence that large ellipical galaxies form the
majority of their stars at early epochs ($z > 1$), and  quickly
(duration $< 1$ Gyr; \citet{Renz06}). Bright submillimeter-selected
galaxies (SMGs; defined as $S_{850\mu m} > 3$\,mJy) are an important galaxy
population in this regard, likely representing a key star formation
epoch with  extreme starbursts thought to be driven by
major mergers of gas-rich galaxies at high redshift \citep{Smai02, Blai02, Tacc08}. The discovery of evolved galaxies
at high redshift \citep[$z > 2$;][]{Dadd05, Krie08, Toft09, vDok08, Kurk09, Coll09, Will09, Koti09}, as well as luminous SMGs at even higher redshifts  \citep[$z > 5$;][]{Riec10, Capa11}
provides evidence for this early active star formation phase  out to about 1 Gyr after the big bang.

The very bright SMG MM18423+5938  ($S_{1200 \mu m}=30 \pm 3$ mJy)
was discovered serendipitously at the IRAM 30m telescope with the
bolometer array MAMBO \citep{Lest09}. 
Although no optical counterpart was known for MM18423+5938,  its redshift
$z=3.92960\pm 0.00013$  was spectroscopically  measured through the detection 
of several strong CO and CI  emission lines
with the  wide spectrometer EMIR also at  this telescope \citep{Lest10}.
These line detections indicate a massive  reservoir of molecular gas in a galaxy only 
1.6 Gyr after the big bang. The observed CO ladder  (CO(4-3), CO(6-5) and CO(7-6), CO(9-8) upper limit) 
peaks around $J=5$, indicative of a starburst-powered SMG that is not dominantly excited by an active nucleus (AGN).  

Recently, bright, strongly lensed SMGs have been discovered in large area surveys \citep{Viei10, Negr10, Fray11}. 
Even compared to these, MM18423+5938 is very bright, and its well established redshift makes it a prime target 
for high-resolution observations to study the cause of its high brightness, to understand its place 
in this emerging population, to verify the hypothesis that it is gravitationally lensed, and to study further
the similarities of MM18423+5938 with the other very bright SMG SMMJ2135-0102 at $z=2.3$ ($S_{850\mu m}$=106~mJy)
which has been shown to be strongly lensed \citep{Swin10}.

\begin{figure*}
\center  
\epsscale{1.05}
\plotone{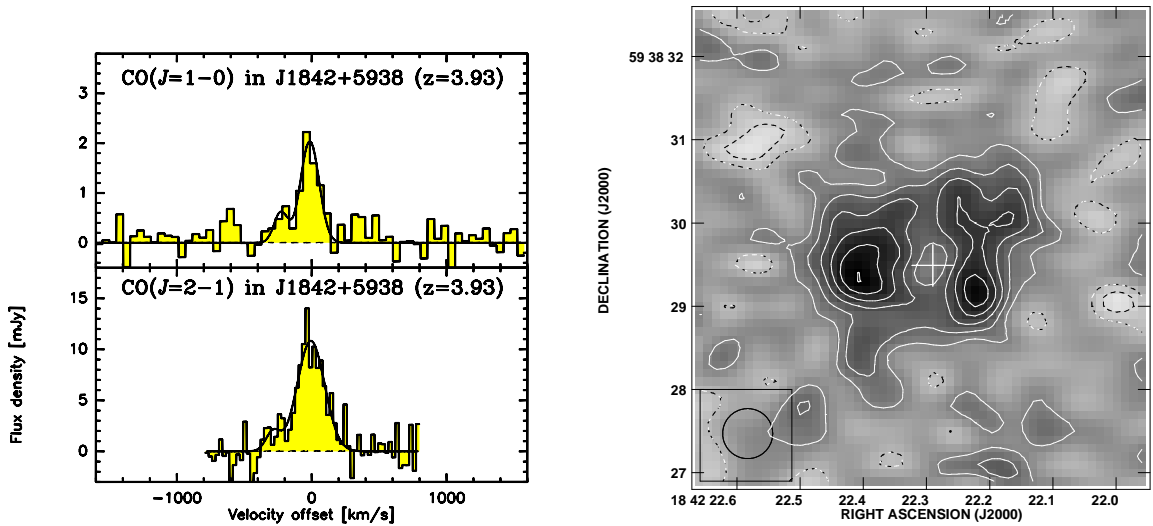}  
\caption{EVLA observations. {\sl Top left} : spectrum of the CO(1-0) emission from MM18423+5938 at
51 km s$^{-1}$ resolution.  {\sl Bottom left}~: spectrum of the  CO(2-1) emission at 26 km s$^{-1}$ resolution. 
Zero velocity corresponds to the redshift $z = 3.9296$. {\sl Right~:}~image 
of the CO(2-1) emission from MM18423+5938 integrated over 230 km~s$^{-1}$ at
$0.6''$ resolution. The rms  is 0.16 mJy beam$^{-1}$, and
the contour levels are $-0.4, -0.2$, 0.2, 0.4, 0.6, 0.8, 
1.0, 1.2 mJy beam$^{-1}$.  Negative contours are dashed. The
cross shows the position of the lensing galaxy at the ring center.
\label{EVLA}}
\end{figure*}

\begin{deluxetable*}{cccccc}
\tablecaption{Observed line parameters of  2-component gaussian fit \label{tab:lines}}
\tablehead{
 line & \colhead{$\rm S_{\nu}$} & \colhead{$\Delta$V$_{\rm FWHM}$} & \colhead{$\mathrm{V_{0}}$} & \colhead{$\mathrm{\rm I_{CO}}$} & \colhead{$L'_{CO} 10^{-11}$} \\
      & \colhead{(mJy)}         & \colhead{(km/s)}                 & \colhead{(km/s)}           & \colhead{(Jy km/s)}             & \colhead{($\rm K~km/s~pc^{2}$)} }  
\startdata
CO(1-0) & $2.05 \pm 0.22$   & $161\pm 30$   & $-14 \pm 11$    & $0.35 \pm 0.05$    & $2.71 \pm 0.38$ \\
        & $0.62 \pm 0.38$   & $118\pm 114$  & $-228 \pm 52$   & $0.078 \pm 0.06$   &                  \\
CO(2-1) & $10.83 \pm 0.74$  & $239\pm 26$   & $-6 \pm 9$      & $2.74 \pm 0.23$    &  $4.74 \pm 0.40$   \\
        & $1.92 \pm 1.34$   & $125\pm 128$  & $-286 \pm 59$   & $0.25 \pm 0.20$    &                   
\enddata
\tablecomments{Zero velocity corresponds to a
redshift  $z = 3.9296$. 
The apparent $L'_{CO}$ results from the two components added up. Only statistical uncertainties are given. }
\end{deluxetable*}

In this Letter we report observations of the CO(1-0) and CO(2-1) line emissions
in MM18423+5938  to $0.6''$ resolution using the Expanded Very
Large Array (EVLA; \citet{Perl11}), as well as multicolor optical and near-IR imaging using the
Canada-France-Hawaii Telescope (CFHT). These observations confirm  that MM18423+5938 is lensed,
 providing a unique opportunity to study the gas dynamics on small physical scale with  further  
high sensitivity and high spatial resolution EVLA observations. 
We use a  concordance, flat $\Lambda$-CDM 
cosmology throughout,  with $H_0$ = 71 km s$^{-1}$ Mpc$^{-1}$, 
$\Omega_{\small M} $ = 0.27, and $\Omega_{\Lambda}$ = 0.73. 
% (\citet{Sper03}, \citet{Sper07}).

%%%%%%%%%%%%%%%%%%%%%%%%%%%%%%%%%%%%%%%

%\section {Observational data and reductions} \label{data}

\section {Observational data and reductions} 

%%%%%%%%%%%%%%%%%%%%%%%%%%%%%%%%%%%%%%%

\subsection{Expanded Very Large Array}

Observations were made in the C configuration of the EVLA in December 2010. We targeted  the CO(1-0) line
(rest frequency = 115.271 GHz) at 23.38 GHz, and the CO(2-1) line (rest
frequency = 230.538 GHz) at 46.76 GHz. A total bandwidth of 248 MHz
was employed, with 124 spectral channels.
Spectral imaging of the broad line of  MM18423+5938 at high spectral resolution was made possible with
the new capability of the  WIDAR correlator of the EVLA \citep{Perl11}. The total observing time
was 2 hours at 23 GHz and 8 hours at 47 GHz. Dynamic scheduling was
employed to ensure good atmospheric observing conditions. The phase calibrator
was J1849+6705 and the switching cycle was 5 minutes \citep{Cari99}.  
The synthesized beam FWHM using robust weighting  (R=1)  in the AIPS data analysis software  
was $2.1"\times 1.0"$ at 23 GHz, and roughly circular with FWHM $=
0.6"$ at 47~GHz.  The flux density scale was set by observing 3C48,
but some observations at 47~GHz were made with the target
source and 3C48 at very different elevations. We estimate an absolute calibration 
uncertainty at 47~GHz of $\pm 20\%$, even with the nominal
opacity correction made using the weather data \citep{Butl10}.
The rms noise per channel is 0.16 mJy beam$^{-1}$ at 23 GHz (rest frequency = 115.271 GHz)  at 51 km
s$^{-1}$ resolution, and 0.48 mJy beam$^{-1}$ at 47~GHz (rest
frequency = 230.538 GHz)  at 26 km s$^{-1}$ resolution.

Figure \ref{EVLA} (left panel) shows the  CO(1-0) and CO(2-1) spectra. 
Both lines show a possible non-Gaussian tail towards lower velocities, so two components
were fitted to both the CO(1-0) and CO(2-1) profiles. 
Line peak flux densities, line widths, mean velocities and velocity-integrated line fluxes $\rm I_{CO}$ 
are in Table  \ref{tab:lines}.

\begin{figure*}[t]
\epsscale{0.80}
\plottwo{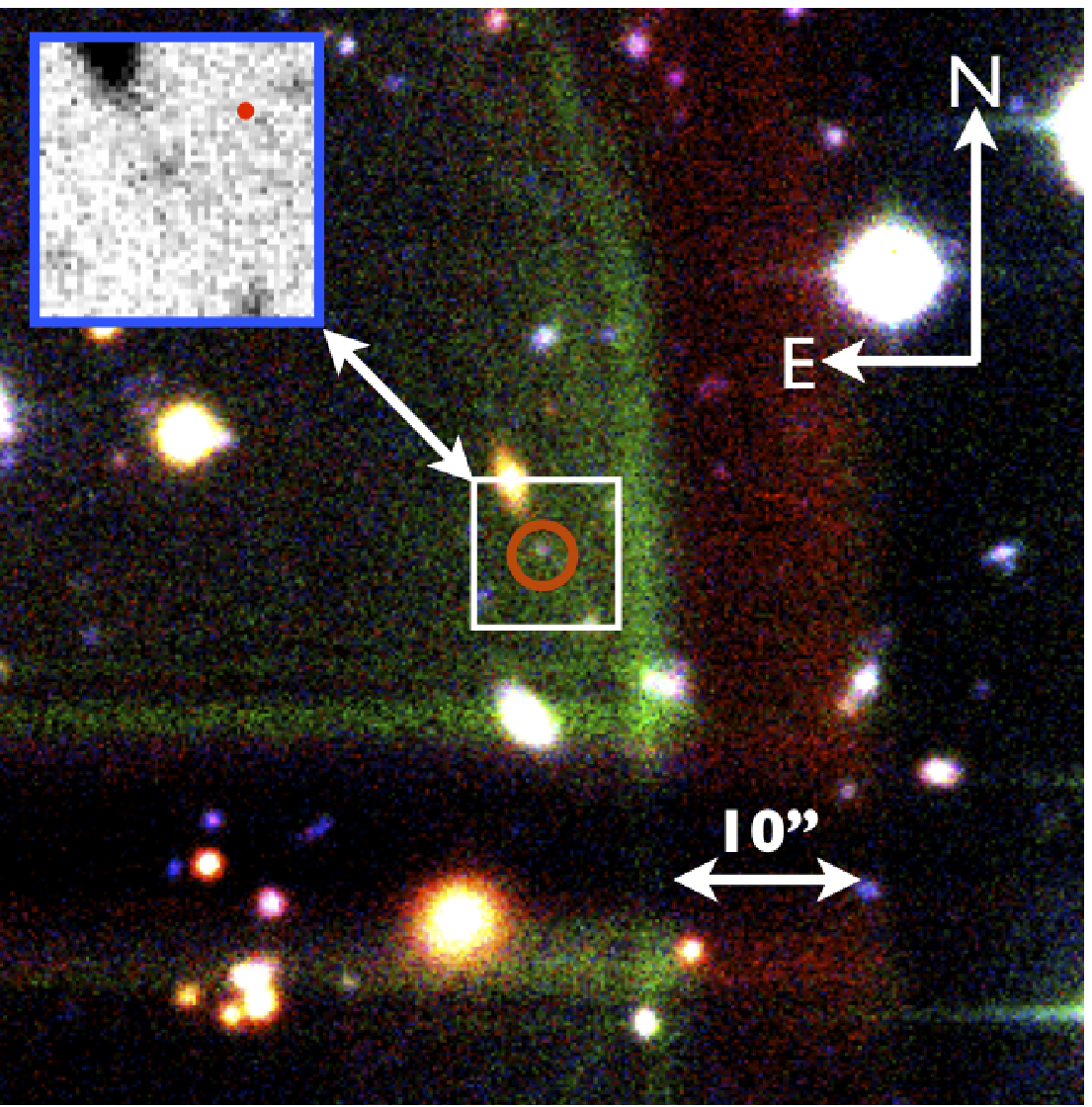}{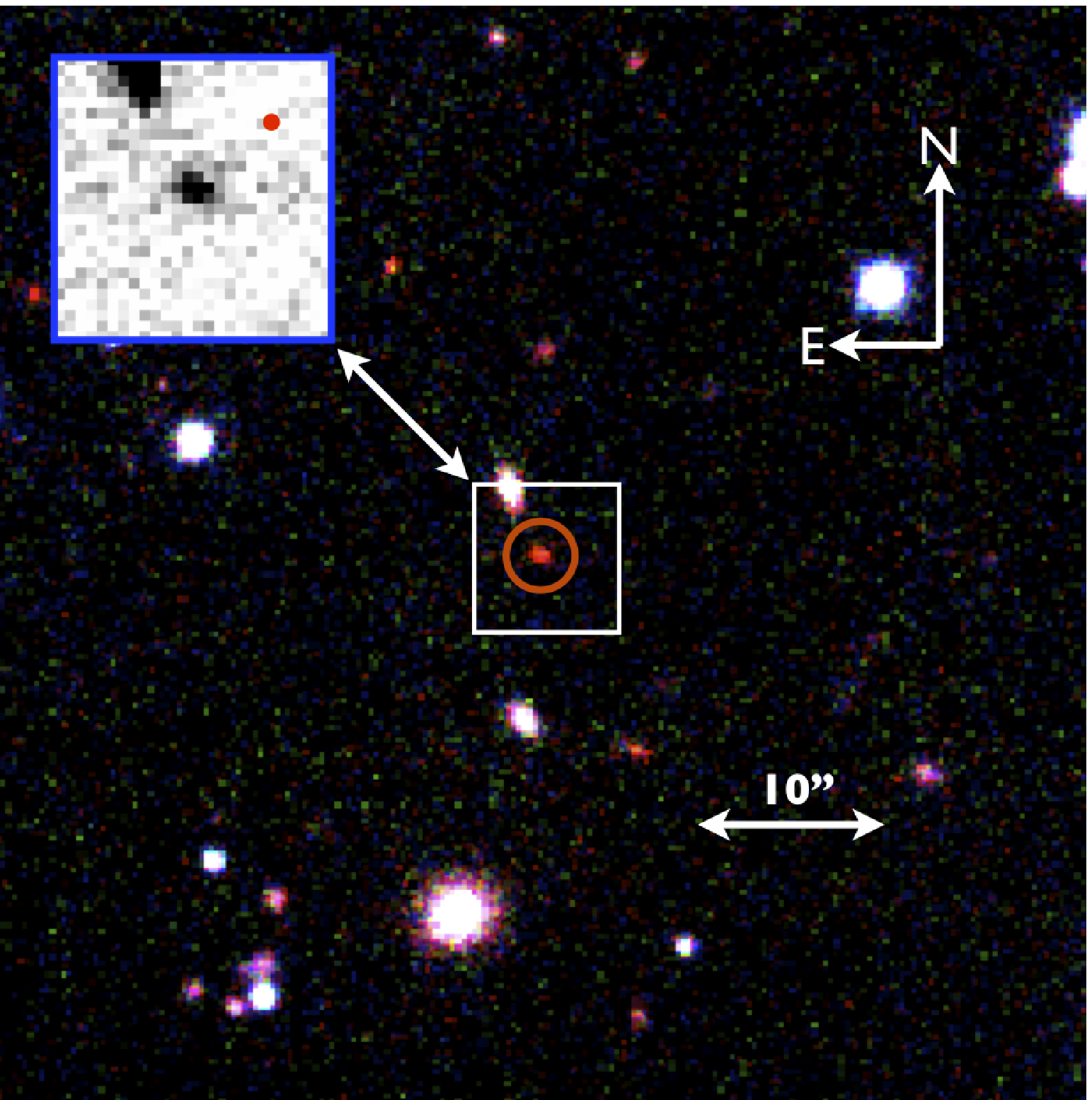}
\caption{CFHT images. RGB color images of $60'' \times 60''$ regions centered on MM18423+5938  in
the stacked MegaCam $g, r,$ and $i$, and WIRCam $J, H$, and $K_{\rm S}$  band images, respectively, are shown in the left and right panels.  
The pixel scales are  0".187 and  0".304 in the  MegaCam and WIRCam images, respectively.
The main lens galaxy is circled in red and is at the center of the insets ($8'' \times 8''$). 
The red dots in the insets are the FWHM point-spread function (PSF) during observations 
(in r-band, and Ks band). The optical image was affected by scattered light from two
nearby $\sim 9$ Vmag stars, see \S 2.2 for details.
\label{CFHT_griRGB}}
\end{figure*}

\begin{deluxetable*}{lccccccc}
\centering
\tablecaption{ CFHT {\sc griz} and JHK$_{\rm S}$ Band Photometry  (Extinction Corrected AB-Magnitudes) \label{tab:CFHT_photo}}
\tablehead{
 Object         &  $g$            & $ r $            & $ i $            & $z$            & $J$             & $H$              & $K_S$   }    
\startdata
 { Lens Galaxy }    &   {$26.33\pm0.43$ } &  { $24.84\pm0.20$ }  & { $24.17\pm0.15$ }  &   ....          &  { $21.81\pm0.24$ }  &   { $21.25\pm0.33$ }  &  { $20.74 \pm0.08$ }   \\

 { Nearby Gal. (N.E.) } &  { $23.75\pm0.05$ } & { $22.28\pm0.03$ }  &   { $21.33\pm0.02$ }  &  { $20.67\pm0.04$ }  &   {$19.19\pm0.03$ } &  { $19.14\pm0.04$ }  &  { $19.08\pm0.02 $ }   
\enddata
%\tablecomments{ footnote can be added }
\end{deluxetable*}

Figure \ref{EVLA} (right panel) shows the CO(2-1) emission integrated over the channels of the full line width
(230 km~s$^{-1}$) at $0.6''$ resolution.  The morphology is consistent with an Einstein ring interpretation. The ring
is elliptical, with a major axis $\sim 1.4"$ and a minor axis $\sim
1.2"$. The J2000 coordinates of the ring center (geometric center of the inner hole)  are  
$\alpha$= 18h~42m~22.29s, $\delta=59^{\circ}~38'~29.5''~\pm0.2''$ in the International Celestial Reference System (ICRS). 
No continuum emission is detected at either frequency, with 3$\sigma$ limits of
75 $\mu$Jy beam$^{-1}$ at 23 GHz and 210 $\mu$Jy beam$^{-1}$ at 47 GHz.

%%%%%%%%%%%%%%%%%%%%%%%%%%%%%%%%%%%%%%%
\subsection {CFHT Optical and Near-IR imaging}  
%%%%%%%%%%%%%%%%%%%%%%%%%%%%%%%%%%%%%%%

To identify the putative lensing galaxy, we carried out multi-band
optical and Near-IR (NIR) imaging at the CFHT with the prime focus imagers, $MegaCam$ and
$WIRCam$ respectively. The observations were completed in 2010 November, 
with total on-source integration times
of 4.75h on MegaCam (6300s in $i$, 5400s in $r$, and 2700s each in $g$ and
$z$ filters), and  0.58h on WIRCam (950s in K$_{\rm S}$, 840s in $J$, and 285s in
$H$bands). The median seeing during the optical imaging was 0".65 in the $r$-band, which
was well sampled by the MegaCam pixel scale of 0".187; during the NIR imaging, the
median seeing was 0".7 in  K$_{\rm S}$-band, marginally sampled by the WIRCam 0".304 pixel scale.
Standard calibrations were obtained and the data processed at CFHT
with the $Elixir$ and $I'iwi\,v.2$ pipelines. The sky-subtracted dithered exposures
in each filter were stacked using $Swarp$ \citep{Bert02} and MegaPipe \citep{Gwyn08}. 
For obtaining consistent photometry in all passbands,
 the NIR images were re-convolved from 0".304
pixel scale to the MegaCam resolution of 0".187 during
stacking of the dithered WIRCam images. Astrometric and
photometric calibrations in the optical and NIR were carried out with $Scamp$
\citep{Bert06a}, against USNO-B1 \citep{Mone03}, and 2MASS Point Source  \citep{Skru06} catalogs.
The 2MASS positions are tied to the ICRS system via the Hipparcos Tycho reference frame to
better than $0.2''$ (2MASS Explanetary Supplement Chap I, subsection 6b, paragraph ix).
There are thirteen  2MASS sources in our full WIRCam image that have been used for this NIR-radio registration. 
The RGB color stamp images (Figures \ref{CFHT_griRGB})
constructed with optical $gri$ filters, and the NIR $JHK_S$ bands, show a red
galaxy (with ellipticity~=~0.19) at $\rm \alpha$=18h~42m~22.27s$\pm0.29''$,
$\delta=59^{\circ}~38'~29.6''\pm0.25''$ in the ICRS, {\i.e.} in the same coordinate system as our EVLA map. 
In the NIR (Fig \ref{CFHT_griRGB}, \textit{right}), the detection significance is at 10$\sigma$ 
in K$_{\rm S}$, and at $\sim 3\sigma$ in $J$
and $H$ bands. In the optical (Fig \ref{CFHT_griRGB}, \textit{left}), the
bright stellar halo from the two nearby  $\sim 9$ Vmag saturated stars led to
significant background noise, and therefore to lower detection significances but nonetheless  
successful identifications in the $i,g,r$ filters. The $z$ filter was affected by fringing.

Given the low detection significances in the optical imaging, magnitudes were
measured using $Sextractor$ \citep{Bert96} in dual image mode, using the K$_{\rm S}$
band image for object detection, while measuring the flux in matched apertures
in all filters; for the optical filters, the WIRCam K$_{\rm S}$ image was resampled
to match the pixel scale of MegaCam. 
 Further, we carried out photometry of the nearby galaxy at $\rm \alpha=18h~42m~22.52s\pm0.36''$, 
$\delta=59^{\circ}~38'~33.17''\pm0.31''$ in the ICRS
(lying $5''$ E of N of the putative lens, see Fig \ref{CFHT_griRGB}) expected to contribute to 
the lensing model. The foreground Galactic extinction corrected \textit{AB}-magnitudes are 
in Table  \ref{tab:CFHT_photo} (extinction $E(B-V)$ is $0.0462\pm 0.0009$ 
at the source position in Schleger maps).

\section{Analysis and discussions} \label{Analysis}

\begin{figure}
\epsscale{1.0}
\plotone{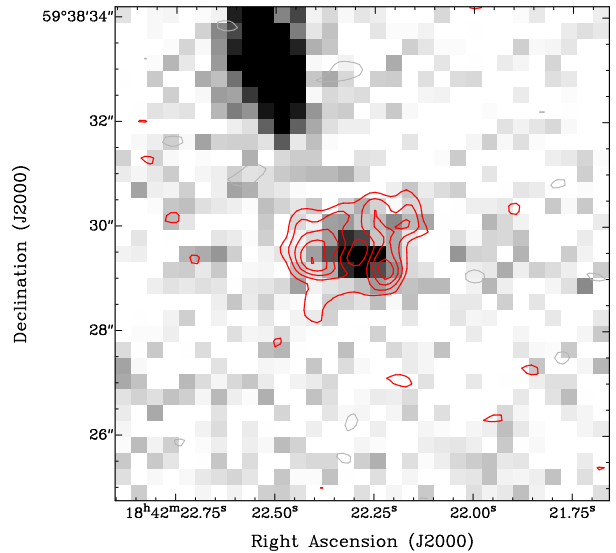}
\caption{Overlay of the CO(2-1) contour map  of
MM18423+5938 and the K$_{\rm S}$ band image of the main lens galaxy and N.E. nearby galaxy (gray scale). Contours
are the same as in Fig \ref{EVLA} (right panel).
Pixel size of the K$_{\rm S}$ band image is $0.304''$.  \label{overlay}}
\end{figure}

Figure \ref{overlay} shows an overlay of the CO(2-1) contour map  
of MM18423+5938  and the K$_{\rm S}$ band image. The 
main lens galaxy  is coincident with the geometric center of the ring.
With the measured NIR and optical magnitudes,
we used SED fitting in \textit{LePhare}  \citep{Arno99} to estimate a
photometric redshift $z_{phot}\sim1.1$, with a formal uncertainty of $\pm 0.1$,
for the lens best fit with an early type galaxy
template. No satisfactory fit was obtained for the North-East nearby
galaxy mainly because the  J magnitude did not match with  the other bands. 
We built a preliminary lens model based on isothermal
 elliptical potentials for the two galaxies  using the {\it LENSTOOL}
 Software (\citet{Knei96}, \citet{Jull07}). This  resulted
 in a quadruple image with the two brightest 
 spots located at the two maxima of the CO(2-1) map, and yielded 
 the total magnification factor $m \sim 12$. The nearby galaxy contributes to the lensing
 model in making the ring slightly ellipitical as observed but its mass and redshift cannot be
separated with our current data. Had we derived its redshift, we could have speculated whether
 or not the J~band excess could be  due to strong  emission of its $H_{\alpha}$ line.   The 
 physical scale probed in the source with the beam of our EVLA map is
 $0.6'' \times ({7.1 \rm kpc}~{\rm arcsec^{-1}}) / \sqrt{m}$, {\it i.e.} ${\rm 1.2~kpc}$ with our
first estimate of the magnification factor $m$.

We revise the far-IR luminosity  $L_{FIR}$ of MM18423+5938 given in  \citet{Lest10} 
since the strong $Spitzer~MIPS$ $70\mu$m  source,  tentatively 
associated with MM18423+5938 then, has now  a position  that is discrepant by more 
than  $5\sigma$ with the more accurate EVLA position newly found for  MM18423+5938.
The SED of the dust emission was modeled by  a single temperature 
 modified black-body function with the standard  power law for the dust opacity, fitted to the  available photometry 
at  $\lambda$= 1.2, 2 and 3~mm given by \citet{Lest10} and a 3-$\sigma$ upper limit of 3 mJy at $70\mu$m.  We   
varied the opacity index  over the range $1.0 < \beta <  2.0$ and the temperature over 
$20 < T_{\rm dust} < 60$~K  that are plausible for SMGs \citep[e.g.][]{Magn10, Ivis10, Hwan10, Chap10}, and found   
the range $2 \times 10^{13} ~m^{-1} < L_{\rm FIR} < 3 \times 10^{14}  ~m^{-1} ~{\rm L_{\odot}}$ with a 95\% confidence
level, where $L_{\rm FIR}$ is mostly the luminosity between $\lambda \sim 70\mu$m and 3mm,  
and $m$ is the magnification factor.
Using the FIR to star formation rate conversion from \citet{Kenn98},  this luminosity corresponds to 
a range in star formation rate of $3300  ~m^{-1}  <  SFR  < 22000  ~m^{-1} ~{\rm {\rm M}_{\odot}} {\rm yr}^{-1}$. 
Recently, \citet{McKe11} made a deep image at 1.4 GHz with the
Westerbork array (WSRT)  and detected MM18423+5938 at a position consistent with
ours to within $1\sigma$.  They used the  radio-FIR relation
to infer an apparent luminosity $L_{FIR}= 5.6^{+4.1}_{-2.4} \times 10^{13} L_{\odot}$
consistent with our new estimate.

The  velocity-integrated CO(1-0) flux density
yields a luminosity ${\rm L'_{CO(1-0)}} = 2.71\pm 0.38 \times 10^{11} ~m^{-1}$ ~K km
s$^{-1}$ pc$^2$ using the relationships from \citet{Solo92}.  
If corrected by $m \sim 12$, this is the typical luminosity found for  SMGs, {\it e.g.}, 
\citet{Harr10},  \citet{Fray11}, \citet{Ivis11}.      
This implies an H$_2$ mass of $2.2 \times 10^{11} ~m^{-1}  $ M$_\odot$ in
assuming a CO(1-0) luminosity to H$_2$ mass conversion factor
of $\alpha~=~0.8$~M$_\odot$~/~K~km~s$^{-1}$~pc$^2$, suitable for
starburst galaxies in the nearby Universe \citep{Down98} 
and typically used for SMGs, {\it e.g.} \citet{Tacc08}. 
The gas-to-dynamical mass ratio exceeds unity for the dynamical mass ($4.8 \times 10^{10} ~m^{-1}$ M$_\odot$)
computed for an hypothetical edge-on disk  by
using the FWHM of the lines ($\sim 200$ km/s) as an indication of the rotation velocity. 
This exceedingly large ratio and the fact that a disk in  MM18423+5938 is not expected   
to be supported by the radiation pressure of an AGN \citep{Scov95} suggest that a disk would be inclined. 
Note that this ratio still exceeds unity if a merger model is used instead, the enclosed dynamical mass is approximately 
a factor of 2 larger in this case \citep{Genz03}.  In Fig~\ref{EVLA}, 
the second component in the CO(1-0) and CO(2-1) lines detected at a low significance level
is blueshifted by ~200-300km s$^{-1}$. We have verified that this component is not visible 
in the averaged  profile of the three high order (J=4,6,7) CO lines  
observed at the IRAM 30m \citep{Lest10}, which indicates there are distinct  CO components
that have low excitation only within MM18423+5938.
Such a component would be  indicative of either a more extended low order CO  
component of the rotating disk or possibly the second component of a merger.

The high order CO excitation ladder observed with  EMIR/IRAM was fitted
with a large velocity gradient (LVG radiative transfer model, 
$T_k=45 K$, $\rm n(H_2)=10^3 cm^{-3}$, $\rm N_{CO}/\Delta V=3 \times 10^{18} cm^{-2}/(km~s^{-1}$))
in Lestrade et al. (2010). The CO(1-0) peak flux
density ($2.1 \pm 0.2$ mJy) is consistent with this model, while
the CO(2-1) measurement ($10.8 \pm 2.2$ mJy) is marginally in excess by $54 \pm 30 \%$ when the 20\% calibration uncertainty 
at 47~GHz is included.  Although this uncertainty is presently large, we note that the ratio
of the flux densities CO(2-1)/CO(1-0) is  difficult to explain even with differential
lensing if the excitation is spatially variable in the galaxy, unless CO(1-0) is moderately optically
thick, with rising optical depth toward CO(2-1).  
New EVLA observations are required to investigate this in more detail.
 
The peak rest-frame  brightness temperature of the ring is 12 K (East clump in our map of Fig~\ref{EVLA}). This is
similar to what has been seen in other $z~\sim~4$~SMGs, such as GN20
\citep{Cari10}, although we cannot rule-out higher $T_B$
clumps that are not resolved by our present observations.

%%%%%%%%%%%%%%%%%%%%%%%%%%%%%%%%%%%%%%%

\section{Conclusions} \label{Conc}

While there is a long history of CO observations from strongly lensed,
high redshift galaxies \citep[e.g.][]{Allo97,Barv94, Swin10}, 
MM18423+5938 is only the second complete Einstein ring seen in CO emission,
after the  z=4.1 quasar host galaxy PSS J2322+1944 \citep{Cari03}.  
We identified the lens and derived a magnification factor of $\sim 12$ using
a preliminary  lensing model. 
We measured a luminosity $\rm L'_{CO(1-0)}$, and corresponding $\rm H_2$ mass, for  MM18423+5938 
that, when corrected for magnification, are similar to the unlensed SMGs. 
However, the relatively large range for $L_{FIR}$ in our present analysis  
requires further observations to better sample the FIR SED in order to better constrain the implied star formation rate. 
As demonstrated by \citet{Riec08} for  
the molecular Einstein ring observed towards the QSO PSS J2322+1944, 
the brighter SMG MM18423+5938  provides a unique opportunity to study  the gas dynamics on physical scale as
small as $\sim 100$ pc with  further  
high sensitivity and high spatial resolution EVLA observations.

%%%%%%%%%%%%%%%%%%%%%%%%%%%%%%%%%%%%%%%

\acknowledgments
\section*{Acknowledgement}

We acknowledge and sincerely thank Dr.~S.~D.~J. Gwyn (CADC, Victoria, Canada)
for promptly stacking the CFHT $Elixir$ processed MegaCam data with
 $MegaPipe$. This work is based on EVLA observations 
conducted under the auspices of NRAO,  WirCam and MegaCam 
observations conducted  at CFHT, 
and on use of data products from  the Two Micron All Sky Survey (2MASS).

% Based on observations obtained with WIRCam, 
% a joint project of CFHT, Taiwan, Korea, Canada, France, at the Canada-France-Hawaii Telescope (CFHT) 
% which is operated by the National Research Council (NRC) of Canada, the Institute National des Sciences 
% de l'Univers of the Centre National de la Recherche Scientifique of France, and the Univsity of Hawaii. 
% Based on observations obtained with MegaPrime/MegaCam, a joint project of CFHT and CEA/DAPNIA, 
% at the Canada-France-Hawaii Telescope (CFHT) which is operated by the National Research Council (NRC) 
% of Canada, the Institut National des Science de l'Univers of the Centre National de la Recherche 
% Scientifique (CNRS) of France, and the University of Hawaii. This publication makes use of data 
% products from the Two Micron All Sky Survey, which is a joint project of the University of 
% Massachusetts and the Infrared Processing and Analysis Center/California Institute of Technology, 
% funded by the National Aeronautics and Space Administration and the National Science Foundation. 
% This research has made use of the USNOFS Image and Catalogue Archive operated by the United States 
% Naval Observatory, Flagstaff Station (http://www.nofs.navy.mil/data/fchpix/).

%%%%%%%%%%%%%%%%%%%%%%%%%%%%%%%%%%%%%%%

{\it Facilities:} \facility{EVLA}, \facility{CFHT}.

%%%%%%%%%%%%%%%%%%%%%%%%%%%%%%%%%%%%%%%

% \appendix

%%%%%%%%%%%%%%%%%%%%%%%%%%%%%%%%%%%%%%%

%\bibliographystyle{apj}
%\bibliography{ms_emulApJ}

%\end{document}

\end{document}